\documentclass[prd,showpcs,amsmath,amssymb,nofootinbib,longbibliography,twocolumn,showpacs,notitlepage]{revtex4-1}
\usepackage{multirow}
\usepackage{epsfig}
\usepackage{amsmath}
\usepackage{bm} 
\usepackage{times}
\usepackage{graphicx}
\usepackage{color}
\usepackage{slashed}
\usepackage{graphicx}
\usepackage{amsmath} 
\usepackage[latin1]{inputenc}
\usepackage{hyperref}
\usepackage{soul}
\usepackage{multirow}
\usepackage{epsfig}
\usepackage{amsmath}
\usepackage{bm} 
\usepackage{times}
\usepackage{graphicx}
\usepackage{color}
\usepackage{slashed}
\usepackage{graphicx}
\usepackage{amsmath}
\usepackage{tikz}
\usetikzlibrary{positioning,shapes}
\usepackage{relsize} 
\usepackage[latin1]{inputenc}

\def\bea{\begin{eqnarray}}
\def\eea{\end{eqnarray}}
\def\bean{\begin{eqnarray*}}
\def\eean{\end{eqnarray*}} 
\def\nn{\nonumber}
\def\beaal{\begin{align}}
\def\eeaal{\end{align}}

\begin{document} 
 
\title{Testing Unification and Dark Matter  with Gravitational Waves}

\author{Bartosz~Fornal}
\affiliation{Department of Chemistry and Physics, Barry University, Miami Shores, Florida 33161, USA\vspace{1mm}}
\author{Kassandra~Garcia\vspace{1mm}}
\affiliation{Department of Chemistry and Physics, Barry University, Miami Shores, Florida 33161, USA\vspace{1mm}}
\author{Erika~Pierre\vspace{1mm}}
\affiliation{Department of Chemistry and Physics, Barry University, Miami Shores, Florida 33161, USA\vspace{1mm}}

\date{\today}

\begin{abstract}
We propose to search for a new type of gravitational wave signature relevant for particle physics models with symmetries  broken at vastly different energy scales. The spectrum contains a characteristic double-peak structure consisting of a sharp peak  from domain walls and a smooth bump  from a first order phase transition in the early Universe. We demonstrate how such a gravitational wave signal  arises in a  new theory unifying baryon number and color into an SU(4) gauge group broken at the multi-TeV scale, and with lepton number promoted to an SU(2) gauge symmetry broken at the multi-EeV scale. The model contains two types of  dark matter particles, explains the observed domination  of matter over antimatter in the  Universe, and accommodates nonzero neutrino masses.
We discuss how future gravitational wave experiments, such as LISA, Big Bang Observer, DECIGO,  Einstein Telescope, and Cosmic Explorer, can be utilized to look for this novel signature.\vspace{7mm}
 \end{abstract}

\maketitle

\section{Introduction}
The Standard Model of elementary particle physics is a gauge theory based on the symmetry group
\bea\label{eeq1}
{\rm SU}(3)_c \times {\rm SU}(2)_L \times {\rm U}(1)_Y \ .
\eea
The electroweak sector of the theory was formulated in the 1960s  \cite{Glashow:1961tr,Higgs:1964pj,Englert:1964et,Weinberg:1967tq,Salam:1968rm}, whereas the quantum chromodynamics part  was constructed  in the 1970s \cite{Fritzsch:1973pi,Gross:1973id,Politzer:1973fx}. Since then, the model has  withstood all experimental tests, culminating in the discovery of the Higgs particle at the Large Hadron Collider in  2012 \cite{CMS:2012qbp,ATLAS:2012yve}. Despite this huge success,  the Standard Model on its own is  not complete, since it does not accommodate: $(a)$ dark matter, $(b)$ matter-antimatter asymmetry of the Universe, and $(c)$ nonzero neutrino masses.

The symmetry structure in Eq.\,(\ref{eeq1}) is rather complex and its origins are not understood. There is no theoretical reason to expect that this exact gauge symmetry persists all the way up to the Planck scale. Indeed, the Standard Model symmetry might be just a low-energy manifestation of a larger gauge group existing at higher energy scales, just like ${\rm U}(1)_{\rm EM}$ is a leftover symmetry from the breaking of ${\rm SU}(2)_L \times {\rm U}(1)_Y$ at the electroweak scale. Additionally, the outstanding open questions  require new fields to be introduced into the theory -- such  new fields would most naturally be part of a gauge structure beyond that of the Standard Model.

Specifically, the evidence for the existence of dark matter  has piled up over the last decades \cite{1970ApJ...159..379R,Boomerang:2000efg,Gavazzi:2007vw}, but the strength of its interactions with the Standard Model remains unknown, with only upper limits set by various experiments (for a review, see \cite{Feng:2010gw}). Although the pessimistic  scenario, in which the dark matter particle is completely decoupled from the Standard Model, is a viable possibility, the hope is that the dark matter  actually constitutes an integral part of a larger particle physics structure involving nonzero interactions with  the known particles. In such a case, there must exist a gauge symmetry describing how the dark matter fits into the entire particle physics picture. This framework could also offer explanations for other pressing problems, such as the origin of the matter-antimatter asymmetry of the Universe or the mechanism behind neutrino masses. 

In this work, we propose a new gauge extension of the Standard Model which accomplishes the above-mentioned goals, adopting elements of the models constructed in  \cite{Fornal:2015boa} and \cite{Fornal:2017owa}. This theory contains two possible GeV-scale candidates for the dark matter particle, while the observed excess of matter over antimatter is explained in an asymmetric dark matter setting, i.e., the dark matter and ordinary matter asymmetries share a common origin  \cite{Nussinov:1985xr,Kaplan:1991ah,Hooper:2004dc,Kaplan:2009ag,Petraki:2013wwa,Zurek:2013wia}. The out-of-equilibrium dynamics needed for a successful baryogenesis/leptogenesis is provided by two first order phase transitions in the early Universe. The energy scales for those phenomena are free parameters, and we consider the scenario in which both of them are high, inaccessible  at the Large Hadron Collider. However, such high symmetry breaking scales make the model ideal for being probed  in gravitational wave experiments.

Indeed, the much needed breakthrough in particle physics might come from the detection of primordial gravitational waves. Thus far, the Laser Interferometer Gravitational Wave
Observatory (LIGO) within the LIGO/Virgo collaboration has discovered signals coming from black hole/neutron star mergers \cite{LIGOScientific:2016aoc}, however, a stochastic gravitational  wave background  from the early Universe is predicted by many models of physics beyond the Standard Model. Although LIGO is currently sensitive to a relatively small parameter space of such models, the reach will be considerably improved with future upgrades, as well as other planned gravitational wave experiments, such as the
Laser Interferometer Space Antenna  (LISA) \cite{Audley:2017drz},  Big Bang Observer (BBO) \cite{Crowder:2005nr}, DECIGO \cite{Kawamura:2011zz}, Einstein Telescope (ET) \cite{Punturo:2010zz} and Cosmic Explorer (CE) \cite{Reitze:2019iox}. The possible sources for this stochastic gravitational wave signal are: first order phase transitions in the early Universe after inflation \cite{Kosowsky:1991ua},  inflation itself \cite{Turner:1996ck}, and topological defects (domain walls \cite{Hiramatsu:2010yz} and cosmic strings \cite{Vachaspati:1984gt,Sakellariadou:1990ne}). In the model we propose, gravitational waves originate from annihilating domain walls and a first order phase transition.

Domain walls are topological defects which are created when a discrete symmetry is spontaneously  broken \cite{Kibble:1976sj}. They exist around boundaries of regions corresponding
to different vacua. Stable domain wall configurations  would lead to cosmological problems, since they would overclose the Universe. However, if the two vacua have different energy densities (this difference is called the potential bias), then domain walls become unstable and annihilate, leading to a stochastic gravitational wave background, apriori measurable today. This  can be realized in many particle physics scenarios, e.g.,  electroweak-scale new physics \cite{Eto:2018hhg,Eto:2018tnk,Chen:2020soj,Battye:2020jeu}, supersymmetry \cite{Kadota:2015dza}, Peccei-Quinn symmetry \cite{Craig:2020bnv,Blasi:2022ayo}, high-scale leptogenesis \cite{Barman:2022yos},  left-right symmetry \cite{Borah:2022wdy,Borboruah:2022eex}, grand unification \cite{Dunsky:2021tih}, or 
thermal inflation \cite{Moroi:2011be}.
A review of domain walls and the resulting gravitational wave signatures are discussed in \cite{Saikawa:2017hiv}.

The other source of a stochastic gravitational wave background are cosmological first order phase transitions, which occur  when the effective potential of the theory develops a minimum at a nonzero field vacuum expectation value separated by a potential barrier from the high temperature minimum at zero field value. The transition process from the high temperature false vacuum to the newly formed true vacuum corresponds to bubbles being nucleated in various points in space, which then expand and fill up the entire Universe. The violent expansion of the bubbles results in sound shock waves in the primordial plasma. This, accompanied by bubble collisions and turbulence, leads to  the emission of gravitational radiation. The literature on the subject is extensive and involves various particle physics theories, e.g., electroweak-scale extensions of the Standard Model
\cite{Grojean:2006bp,Vaskonen:2016yiu,Dorsch:2016nrg,Bernon:2017jgv,Baldes:2018nel,Chala:2018ari,Alves:2018jsw,Han:2020ekm,Benincasa:2022elt}, dark gauge groups \cite{Schwaller:2015tja,Breitbach:2018ddu,Croon:2018erz,Hall:2019ank}, dark matter \cite{Baldes:2017rcu,Fornal:2022qim,Kierkla:2022odc}, axions \cite{Dev:2019njv,VonHarling:2019rgb,DelleRose:2019pgi,ZambujalFerreira:2021cte}, 
grand unification \cite{Croon:2018kqn,Huang:2020bbe,Okada:2020vvb}, conformal invariance \cite{Ellis:2020nnr,Kawana:2022fum}, supersymmetry \cite{Craig:2020jfv,Fornal:2021ovz}, left-right symmetry \cite{Brdar:2019fur,Graf:2021xku,Borboruah:2022eex},
seesaw mechanism \cite{Brdar:2018num,Okada:2018xdh,DiBari:2021dri,Zhou:2022mlz}, baryon/lepton  number violation \cite{Hasegawa:2019amx,Fornal:2020esl}), new flavor physics \cite{Greljo:2019xan,Fornal:2020ngq}, and  leptogenesis \cite{Dasgupta:2022isg}. If a phase transition is strongly supercooled, the signal can already be searched for in LIGO/Virgo/KAGRA data sets \cite{LIGO_FOPT}. For a review of gravitational waves from first order phase transitions see \cite{Caldwell:2022qsj}.

In the model we construct, there are two gauge symmetries that are broken at vastly different energy scales. Due to the shape of the effective potential, the first order phase transition happening at a high energy scale ($\sim 100 \ \rm EeV$) leads to the creation of domain walls, whose subsequent annihilation produces a stochastic gravitational wave background peaked in the frequency range relevant for the Einstein Telescope and Cosmic Explorer. The phase transition happening at a lower energy scale ($\sim 100 \ \rm TeV$) is also first order and results in a gravitational wave signal within the sensitivity range of LISA, Big Bang Observer and DECIGO.  
The theory is unique since there are two dark matter candidates and the  matter-antimatter asymmetry can be produced in an asymmetric dark matter framework both at the high and low symmetry breaking scales. Because of this, the predictions for the dark  matter mass derived in \cite{Fornal:2015boa} and \cite{Fornal:2017owa} can be relaxed, possibly leading to intriguing connections to nuclear physics.

We begin by formulating the model in Section \ref{model}, including the symmetry breaking pattern, particle content, masses, and couplings. Then, in Sections \ref{secDM} and  \ref{s3} we discuss the dark matter candidates, as well as the mechanism for leptogenesis at the high scale and baryogenesis at a lower scale. This is followed by a derivation of the expected stochastic gravitational wave signal from domain walls (Section \ref{s4}) and the first order phase transition (Section \ref{s5}). The novel signature involving a combination of the  two signals is discussed in Section \ref{s6}, and followed by conclusions in Section \ref{s7}.

\section{Model} \label{model}

To illustrate the novel gravitational wave signature, we consider a new model combining the features of theories proposed in \cite{Fornal:2015boa} and \cite{Fornal:2017owa}. The model
is based on the gauge symmetry
\bea\label{group}
{\rm SU}(4) \times {\rm SU}(2)_L \times {\rm U}(1)_X \times {\rm SU}(2)_\ell \ ,
\eea
where the $\rm SU(4)$ group unifies color with baryon number, $\rm SU(2)_\ell$ corresponds to  generalized lepton number, 
and $X$ is a linear combination of the diagonal generator of $\rm SU(4)$ and hypercharge. Below we discuss the symmetry breaking  pattern and the particle content of the model, along with the particle masses and couplings relevant for the subsequent analysis of the gravitational wave signal.

\subsection{Symmetry breaking}

The model exhibits a two-step symmetry breaking pattern. We assume that the $\rm SU(2)_\ell$ group is broken first at a very high scale $\sim 10^8 \ {\rm TeV}$, with a subsequent breaking of $\rm SU(4)$ at a lower scale $\sim 100 \ {\rm TeV}$ down to the Standard Model:
 \begin{equation}
     \begin{array}{c}
{\rm SU}(4) \times {\rm SU}(2)_L \times {\rm U}(1)_X \times {\rm SU}(2)_\ell\nn\\[3pt]
\hspace{11.5mm}\bigg\downarrow \hspace{1mm}{\scriptstyle \sim \ 100 \ {\rm  EeV}}\nn\\[12pt]
{\rm SU}(4) \times {\rm SU}(2)_L \times {\rm U}(1)_X\nn\\[3pt]
\hspace{11.5mm}\bigg\downarrow \hspace{1mm}{\scriptstyle  \sim \ 100 \ {\rm  TeV}}\nn\\[12pt]
 \ \ {\rm SU}(3)_c \times {\rm SU}(2)_L \times {\rm U}(1)_Y \ .\\[2pt]
     \end{array}
   \end{equation}   
This choice for the symmetry breaking scales is very different than previously considered  in the literature \cite{Fornal:2017owa,Fornal:2022qim,Fornal:2015boa}.

To implement the above symmetry breaking pattern, as in  \cite{Fornal:2017owa} we introduce two $\rm SU(2)_\ell$ doublet scalars, denoted here by $\hat\Psi_1$ and  $\hat\Psi_2$, governed by the tree-level potential
\bea\label{sc}
&&\hspace{-6mm}V_\Psi(\hat\Psi_1,\hat\Psi_2) = - m_1^2 |\hat\Psi_1|^2 \!-  m_2^2 |\hat\Psi_2|^2 \!- (m_{12}^2 \hat\Psi_1^\dagger \hat\Psi_2+{\rm h.c.})\nn\\
&+& \lambda_1 |\hat\Psi_1|^4 +  \lambda_2 |\hat\Psi_2|^4 + \lambda_3|\hat\Psi_1|^2 |\hat\Psi_2|^2 +  \lambda_4 |\hat\Psi_1^\dagger \hat\Psi_2|^2 \nn\\
&+&  \big[\big(\tilde\lambda_5 |\hat\Psi_1|^2  + \tilde\lambda_6|\hat\Psi_2|^2 +\tilde\lambda_7\hat\Psi_1^\dagger \hat\Psi_2\big)\hat\Psi_1^\dagger \hat\Psi_2+{\rm h.c.}\big] \ . \ 
\eea
Those  scalars develop vacuum expectation values
\bea
\langle \hat\Psi_i \rangle = 
\tfrac{1}{\sqrt2}\,\left(\!
  \begin{array}{c}
    0  \\
    v_i
  \end{array}\!
\right)
\eea
breaking the $\rm SU(2)_\ell$ group and reducing the symmetry  to ${\rm SU}(4) \times {\rm SU}(2)_L \times {\rm U}(1)_X$. It is convenient to define 
\bea
v_\Psi \equiv \sqrt{v_1^2 + v_2^2}\ .
\eea
There are five physical scalar components of $\hat\Psi_1$ and $\hat\Psi_2$, and three gauge bosons from  $\rm SU(2)_\ell$ breaking. 

The second stage of symmetry breaking occurs when the $\rm SU(4)$ quadruplet scalar $\hat\Phi = (4,1,\frac12,1)$, subject to the tree-level potential
\bea\label{sc2}
&&\hspace{-7mm}V_\Phi(\hat\Phi) = -m_\Phi^2 |\hat\Phi|^2 + \lambda_\Phi |\hat\Phi|^4\ ,
\eea
develops the vacuum expectation value \cite{Fornal:2015boa}
\bea
\langle \hat\Phi \rangle = 
\tfrac{1}{\sqrt2}\,(\!
  \begin{array}{cccc}
    0 &
     0 &
      0 &
    v_\Phi
  \end{array}\!
)^T \ .
\eea
This breaks the symmetry down to the Standard Model gauge group, with hypercharge emerging as a combination of $X$  and the diagonal $\rm SU(4)$ generator $T^{15}$,
\bea
Y = X +\tfrac16 \ {\rm diag}(1,1,1,-3) \ .
\eea
The $\rm SU(4)$ breaking leads to seven massive gauge bosons -- six of them form three complex vector fields ${G'}$ transforming as color triplets, and one is the neutral gauge boson $Z'$.

Along with the condition $v_\Psi \gg v_\Phi$, we assume that the coefficients of the cross terms between the $\hat\Psi_i$ and $\hat\Phi$ fields in the scalar potential are small, thus
the two symmetry breaking phenomena can be considered independently of each other.

\subsection{Fermionic particle content}

Given the  structure of the theory, the  Standard Model quarks $Q_L$, $u_R$ and $d_R$ are singlets under $\rm SU(2)_\ell$, but they constitute part of $\rm SU(4)$ quadruplets,
\bea
\hat{Q}_{L} &\equiv& \left(\!
  \begin{array}{cccc}
    Q^r_{L} &
    Q^b_{L} &
    Q^g_{L} &
    \tilde{Q}_{L} 
  \end{array}\!
\right)^T  = (4,2,0,1)\ ,\nn\\
\hat{u}_{R} &\equiv& \left(\!
  \begin{array}{cccc}
    u^r_{R} &
    u^b_{R} &
    u^g_{R} &
    \tilde{u}_{R} 
  \end{array}\!
\right)^T  = \left(4,1,\tfrac12,1\right)\ ,\nn\\
\hat{d}_{R} &\equiv& \left(\!
  \begin{array}{cccc}
    d^r_{R} &
    d^b_{R} &
    d^g_{R} &
    \tilde{d}_{R} 
  \end{array}\!
\right)^T  = \left(4,1,-\tfrac12,1\right)\ .
\eea
The Standard Model leptons $l_L$ and $e_R$, and the right-handed neutrino $\nu_R$ (leading to Dirac masses for the neutrinos), on the other hand, are singlets under  $\rm SU(4)$, but they are part  of  $\rm SU(2)_\ell$ doublets,
\bea
 \hat{l}_L&\equiv& (l_L  \ \ \  \tilde{l}_L)^T = (1,2,-\tfrac12,2) \ , \nn\\
\hat{e}_R &\equiv& (e_R  \ \  \tilde{e}_R)^T =  (1,1,-1,2) \ ,  \nn\\
\hat{\nu}_R &\equiv& (\nu_R  \ \  \tilde{\nu}_R)^T =  (1,1,0,2)  \ .
\eea
To cancel the resulting gauge anomalies, we introduce the same  $\rm SU(2)_\ell$ and  $\rm SU(4)$ singlet fields as in   \cite{Fornal:2015boa}  and \cite{Fornal:2017owa}, i.e., $Q'_R$, $u'_L$, $d'_L$, $l'_R$, $e'_L$,  $\nu'_L$, for each generation separately.

\subsection{Particle masses and couplings}
\label{subb}

These $\rm SU(2)_\ell$ and  $\rm SU(4)$ singlet fields allow all beyond-Standard Model fermions to have vector-like masses,
\bea\label{Yuk}
\mathcal{L}_{Y} \! &=& \!  \sum_i\Big( Y_{l}^{ab}\,\bar{\hat{l}}_L^a \, \hat{\Psi}_i\,  {l'}_{\!\!R}^{b}+ Y_{e}^{ab} \, \bar{\hat{e}}_R^a \, \hat{\Psi}_i\, {e'}_{\!\!L}^{b}
    + Y_{\nu}^{ab}\,\bar{\hat{\nu}}_R^a \, \hat{\Psi}_i\, {\nu'}_{\!\!L}^{b}\Big)  \nn\\[-2pt]
 &&\hspace{-8mm}+ \ Y_{Q}^{ab}\,\bar{\hat{Q}}_L^a \, \hat{\Phi}\,  {Q'}_{\!\!R}^{b}+ Y_{u}^{ab} \, \bar{\hat{u}}_R^a \, \hat{\Phi}\, {u'}_{\!\!L}^{b}
    + Y_{d}^{ab}\,\bar{\hat{d}}_R^a \, \hat{\Phi}\, {d'}_{\!\!L}^{b} + {\rm h.c.} \ .\nn\\
\eea
For order one Yukawas, this results in $\sim 100 \ \rm EeV$ masses for fermions coupling to $\hat\Psi_i$, and $\sim 100 \ \rm TeV$  for those interacting with $\hat\Phi$. In order to have viable asymmetric dark matter candidates, as will be discussed in Section \ref{s3}, some of the Yukawas need to be small, leading to $\sim \rm GeV$ dark matter masses.

The mass matrices for the components of $\hat\Psi_1$ and $\hat\Psi_2$ (i.e., $P_1$, $P_2$, $A$, $C_1$, $C_2$) were derived in \cite{Fornal:2022qim}, and result in large masses for all the scalars except for $A$, whose mass is fine-tuned to be small. The $\rm SU(2)_\ell$ gauge bosons $W'_{1,2}$ and $Z'_\ell$ also develop $\sim 100 \ \rm EeV$ masses. We do not provide detailed formulas here, since for the case of $\rm SU(2)_\ell$ breaking  we are interested only in the gravitational wave signal from domain walls, which is determined solely by the vacuum expectation value $v_\Psi$ and the $\mathcal{Z}_2$ symmetry breaking parameter $m_{12}^2$.

The masses of the gauge bosons $G'$ and $Z'$ arising from  $\rm SU(4)$ symmetry breaking are
\bea
m_{G'} &=& \tfrac12 g_4v_\Phi\ , \nn\\
m_{Z'} &=& \tfrac12 \sqrt{\tfrac32 g_4^2 + g_X^2} \, v_\Phi \ ,
\eea
while the mass of the radial mode of the  scalar $\hat\Phi$ is
\bea
m_{\Phi} = \sqrt{2 \lambda_\Phi} \, v_\Phi \ .
\eea

Since the $\rm SU(4)$ symmetry is broken down to $\rm SU(3)_c$, the gauge coupling $g_4$ needs to match the strong coupling $g_s$ at the symmetry breaking scale. To determine the corresponding value, we run $g_s$  via the renormalization group equation
\bea
\frac{\partial g_{s}(\mu)}{\partial \log\mu} = -\frac{7g^3_{s}(\mu)}{16\pi^2} \ .
\eea
For instance, this gives $g_4(100 \ \rm TeV)  \simeq 0.88$. The value of $g_X$ at that scale is then fixed by the relation \cite{Fornal:2015boa}
\bea
g_X = \frac{g_4 g_Y}{\sqrt{g_4^2 - \frac{2}{3}g_Y^2}} \ ,
\eea
where $g_Y$ is the weak hypercharge coupling at that scale, which leads to $g_X(100  \ \rm TeV) \simeq 0.30$.

\subsection{Non-perturbative interactions}
At energies above the ${\rm SU}(2)_\ell$ breaking scale the model exhibits  non-perturbative dynamics generated by ${\rm SU}(2)_\ell$ instantons \cite{Fornal:2017owa}. Those instantons induce dimension-six interactions of the following form,
\bea\label{in6}
\mathcal{O} \sim \epsilon_{ij}  &&\hspace{-1.5mm}\Big[(l_L^i \cdot \bar{{\nu}}_R)(l_L^j \cdot \bar{{e}}_R) - (\tilde{l}_L^i \cdot \bar{\tilde{\nu}}_R)({l}_L^j \cdot \bar{{e}}_R) \nn\\
&\hspace{-2.5mm}+& (l_L^i \cdot \tilde{l}_L^j)(\bar{\nu}_R \cdot \bar{\tilde{e}}_R)- (l_L^i \cdot \tilde{l}_L^j)(\bar{\tilde{\nu}}_R \cdot \bar{{e}}_R) \nn\\
& \hspace{-2.5mm}+&  (\tilde{l}_L^i \cdot \bar{\tilde{\nu}}_R)(\tilde{l}_L^j \cdot \bar{\tilde{e}}_R) - (l_L^i \cdot \bar{{\nu}}_R)(\tilde{l}_L^j \cdot \bar{\tilde{e}}_R)\Big], \ \ \ \ \ 
\eea
where the dot represents Lorentz contraction and there is an implicit sum over the family indices. Those non-perturbative processes violate the otherwise accidentally conserved Standard Model lepton number. For example, the second term in Eq.\,(\ref{in6}) leads to the process
\bea
\nu_L \,\tilde{e}_L \to \tilde{\nu}_R \,e_R \ ,
\eea
which violates conventional lepton number, $\Delta L = -1$, since out of the four fields participating in this interaction only $\tilde{\nu}_R$ does not carry lepton number (see \cite{Fornal:2017owa} for details). This field is the right-handed component of one of the two dark matter candidates in the model, so the process violates not only  lepton number, but also dark matter number.

\section{Dark matter}\label{secDM}

There are two particles which are singlets under all gauge groups of the theory -- we denote them by $\tilde\nu'_1$ and $\tilde{u}'_1$. They correspond to the following left- and right-handed components of the fields existing before symmetry breaking, 
\bea
\tilde{\nu}'_1: \ \ \ \ \ \ (\tilde\nu'_1)_L \approx {\nu'}^1_{\!\!L} \ , \ \ \ \  (\tilde\nu'_1)_R \approx \tilde\nu^1_R \ ,\nn\\
\tilde{u}'_1: \ \ \ \ \ \ (\tilde{u}'_1)_L \approx {u'}^1_{\!\!L} \ , \ \ \ \  (\tilde{u}'_1)_R \approx \tilde{u}^1_R \ ,
\eea
where we assumed that the dark matter belongs to the  first generation of the extra fermions, which is an arbitrary choice.

\begin{figure}[t!]
\includegraphics[trim={0cm 6.5cm 0cm 6.5cm},clip,width=8cm]{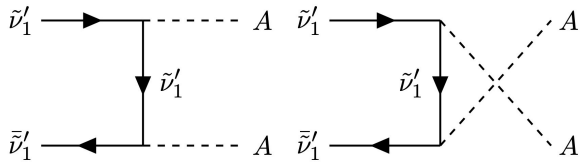} \vspace{-2mm}
\caption{Annihilation channels for the dark matter particle $\tilde{\nu}'_1$.}\label{fig:1}
\end{figure}

The particle $\tilde\nu'_1$ is one of the twelve fermionic states which arise after  ${\rm SU}(2)_\ell$ breaking and develop vector-like masses through the vacuum expectation values of $\hat\Psi_1$ and $\hat\Psi_2$. Due to the conservation of a remnant  ${\rm U(1)}$ global symmetry \cite{Fornal:2017owa} those fermions cannot decay exclusively  to Standard Model particles. Therefore, the lightest of them remains stable and, if it is also electrically neutral, such as $\tilde\nu'_1$, it becomes a good dark matter candidate. The Yukawa matrices can be chosen  such  that the non-dark matter fermions are heavy, whereas ${\tilde\nu'_1}$ is light. 
As demonstrated  in \cite{Fornal:2017owa}, the dark matter annihilation channel to lighter $CP$-odd scalars $A$ (see, Fig.\,\ref{fig:1}),
\bea
{\tilde{\nu}'_1} \  {\bar{\tilde{\nu}}}'_1 \to A \ A \ ,
\eea
is  sufficiently  efficient  to remove the symmetric component of dark  matter. The remaining asymmetric component contributes to the observed dark matter relic abundance, as will be discussed in Section \ref{s3}.

The  breaking of the ${\rm SU}(4)$ symmetry  also results in twelve fermionic states with vector-like masses, this time generated by the vacuum expectation value of $\hat\Phi$. Once again, because of the conservation of an accidental global ${\rm U}(1)$ symmetry \cite{Fornal:2015boa}, their decay channels cannot involve solely  Standard Model particles in the final state. If the lightest of them is the electrically neutral  $\tilde{u}'_1$, it remains stable and becomes another viable component of dark matter. The annihilation channels leading to the correct relic abundance were discussed in \cite{Fornal:2015boa}, and in 
this case a successful  asymmetric dark matter scenario can also be realized.

It is worth emphasizing that with two asymmetric dark matter candidates, their individual masses are not fixed by a single relic density requirement. Depending on the contribution of each of them to the relic abundance, one of them can have a mass smaller than the mass of the neutron, possibly introducing a connection to the dark matter models relevant for the neutron lifetime anomaly \cite{Fornal:2018eol}.

\section{Matter-antimatter asymmetry}\label{s3}

The generation of a matter-antimatter imbalance  in the model proceeds via two independent processes: leptogenesis and  baryogenesis, both occurring in an asymmetric dark matter setting \cite{Nussinov:1985xr,Kaplan:1991ah,Hooper:2004dc,Kaplan:2009ag,Petraki:2013wwa,Zurek:2013wia}. 
 Leptogenesis, as demonstrated in \cite{Fornal:2017owa}, is realized through instantons during ${\rm SU}(2)_\ell$ breaking, whereas baryogenesis, as argued in \cite{Fornal:2015boa}, is achieved within the ${\rm SU}(4)$ sector through the effects of higher-dimensional operators. We discuss both mechanisms below.

\begin{figure}[t!]
\includegraphics[trim={0cm 0cm 0cm 0cm},clip,width=7cm]{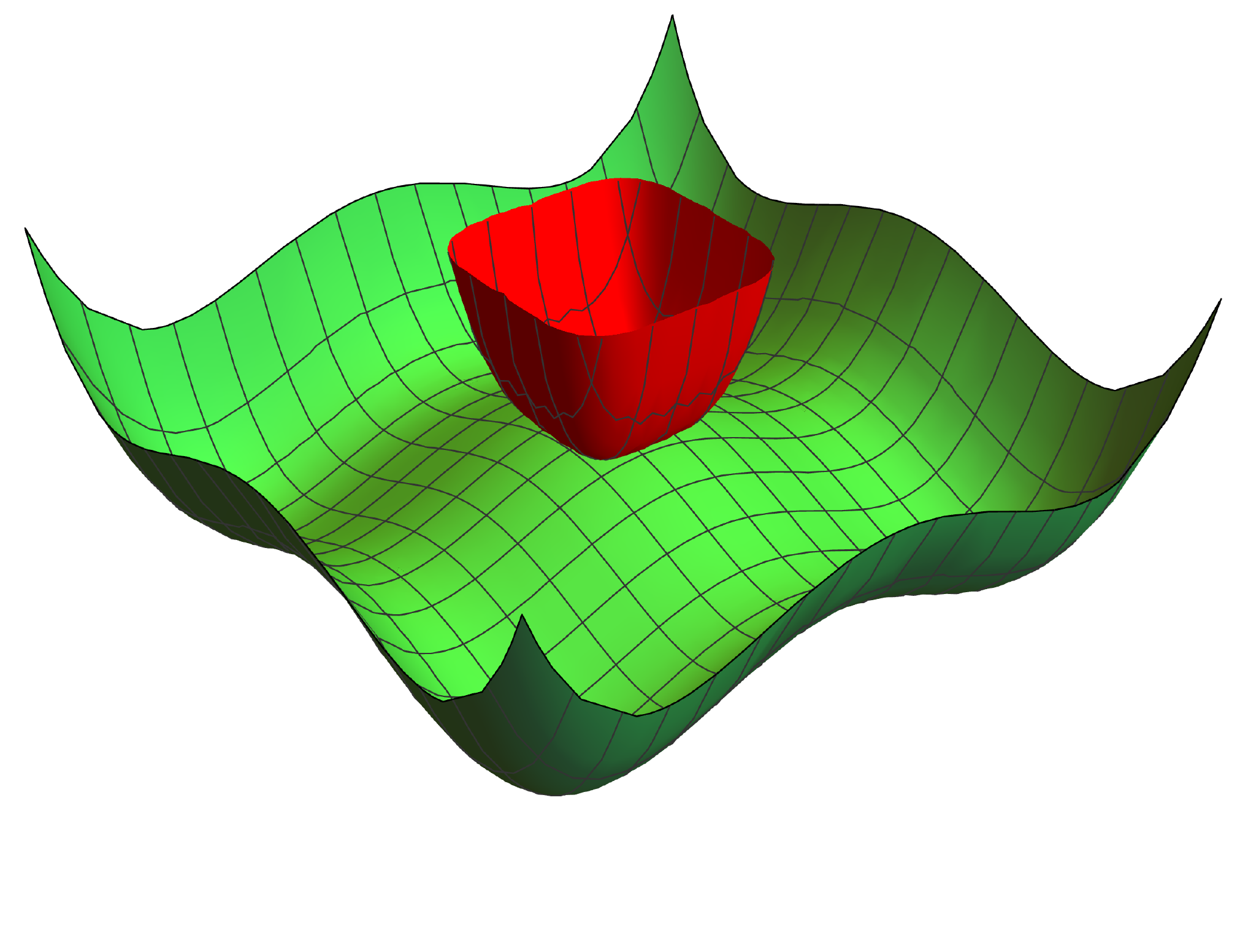} \vspace{-9mm}
\caption{Effective potential of the model $V_{\rm eff}(\psi_1, \psi_2, T)$ at low temperature (green) and high temperature (red).}\label{fig:2}
\end{figure}

The  complete formula for the  effective potential generated by $\hat\Psi_1$ and $\hat\Psi_2$ was derived in \cite{Fornal:2022qim}, where is was demonstrated that a first order phase transition from ${\rm SU}(2)_\ell$ breaking can occur. This scenario is  realized in the model considered in this work, but happens at a higher  symmetry breaking scale. Figure \ref{fig:2} shows a plot of the effective potential $V_{\rm eff}(\psi_1,\psi_2,T)$ produced using {\fontfamily{cmtt}\selectfont
Mathematica} \cite{Wolfram}, assuming the parameters $\lambda_i =0.001$, $v_1=v_2$, $v_\Psi = 100 \ \rm EeV$, $g_\ell=1$, at high temperature (red) and at low temperature (green). 
As the temperature drops, the potential develops new vacua with lower energy densities (see \cite{Fornal:2022qim} for details), separated by a barrier from the high-temperature vacuum. This leads to a first order phase transition, which is precisely the out-of-equilibrium condition enabling leptogenesis to happen, and combines the framework of asymmetric dark matter with  several other leptogenesis/baryogenesis mechanisms \cite{Dick:1999je,Murayama:2002je,Shu:2006mm,Blennow:2010qp}. 

There are four vacua of this type for the scalar potential $V_{\rm eff}(\psi_1, \psi_2)$, but since two of them are related through the gauge transformation $\Psi_i \to e^{i\theta} \Psi_i$, only two of the four vacua are physically distinct \cite{Ginzburg:2004vp,Battye:2011jj}. We denote them by  $\vec\psi_{\rm vac1}$ and $\vec\psi_{\rm vac2}$. The degeneracy between the energy densities of these vacua is broken by the $\mathcal{Z}_2$ symmetry breaking terms in the effective potential, primarily governed by the parameter $m_{12}^2$.

The first order phase transition  corresponds to  nucleation of bubbles with true vacuum inside. Outside the bubble the non-perturbative ${\rm SU}(2)_\ell$ instanton-induced processes described by Eq.\,(\ref{in6}) remain active. This, along with a sufficiently large  amount of $CP$ violation (provided by the appropriate terms in the scalar potential for  $\hat\Psi_1$ and $\hat\Psi_2$),  leads to the generation of a lepton number excess. As the bubble expands, those regions get trapped inside the bubble, where lepton number violation no longer occurs, leading to the accumulation of lepton number  in the Universe. Since the instantons also violate the global dark matter ${\rm U}(1)$ symmetry \cite{Fornal:2017owa}, this results in the generation of a dark matter asymmetry  as well. This process is described by a set of twelve diffusion equations, which have the general form  \cite{Joyce:1994zn,Cohen:1994ss}
\bea
\frac{\partial{\rho_{Ni}}}{\partial t} = D_i \nabla^2 \rho_{Ni} - \sum_j \Gamma_{ij} \frac{\rho_{Nj}}{n_j} + \gamma_i \ .
\eea
In the expression above, $\rho_{Ni}$ is the particle number density, $D_i$ and $\Gamma_{ij}$ are the diffusion constant and  rate, respectively, $n_j$ is the number of degrees of freedom, and $\gamma_i$ is the $i$-th $CP$-violating source given by \cite{Riotto:1995hh}
\bea
\gamma_i \approx \frac{\lambda_7 \mu_{12}^2}{32\pi}\frac{\Gamma_{\psi_i} T_*}{m_{\psi_i}^3\!(T_*)} \partial_z \psi_i \ ,
\eea
where $T_*$ is the bubble nucleation temperature,  $\Gamma_{\psi_i}$ is the decay rate of $\psi_i$, the $z$-axis is in the direction perpendicular to the bubble wall, and $\mu_{12}^2 = {\rm Re}(m_{12}^2)$.
Solving the diffusion equations for our set of parameters reveals that the  generated lepton asymmetry versus the dark matter asymmetry is
\bea\label{LL}
\frac{|\Delta L|}{|\Delta \tilde{\nu}'_1 |} = 3 \ ,
\eea
consistent with the existing result \cite{Fornal:2017owa}. 
The amount of lepton asymmetry produced in this process is subsequently altered by the electroweak sphalerons \cite{Harvey:1990qw}, which partially convert it into a baryon asymmetry,
\bea\label{Baryon}
|\Delta B| = \frac{84}{79}|\Delta  \tilde{\nu}'_1| \ .
\eea
Assuming a bubble wall velocity equal to the speed of light, we find that for $v_\Psi =  100 \ {\rm EeV}$ and  $\lambda_i \sim 10^{-4}$  the  observed baryon-to-photon  ratio \cite{ParticleDataGroup:2022pth}
\bea
\eta \sim  6 \times 10^{-10} 
\eea
is generated provided that
\bea\label{condition2}
|\lambda_7 \mu_{12}^2|\,Y^2 \sim 10^{-10} \  {\rm EeV^2} \ .
\eea
This condition can be satisfied by a wide range of parameter values, e.g., $\lambda_7 \sim 10^{-6}$, $\mu_{12}^2 \sim 10^{-2} \, {\rm EeV^2}$, and $Y \sim 10^{-1}$. The dark matter mass is then given by
\bea\label{m1}
m_{ \tilde{\nu}'_1} \approx \frac{\Omega_{ \tilde{\nu}'_1}}{\Omega_B} \frac{|\Delta B|}{|\Delta \tilde{\nu}'_1 |} \,m_{\rm proton} \ .
\eea
If the $ \tilde{\nu}'_1$ particle makes up all of the dark matter in the Universe, this condition implies $m_{ \tilde{\nu}'_1} \approx 5 \ {\rm GeV}$. However, $\tilde{\nu}'_1$ can be lighter  if the dark matter in the Universe consists also of the $\tilde{u}'_1$ particles. 

Indeed, an analogous asymmetric dark matter mechanism  can generate a contribution to the baryon and dark  matter asymmetries in the $\rm SU(4)$ sector. As pointed out in \cite{Fornal:2015boa}, the following  dimension-six operators,
\bea
\frac{c_1}{\Lambda^2} \epsilon_{abcd} \,\hat{u}^a_R \hat{u}^b_R \hat{d}^c_R \hat{d}^d_R\ , \  \ \ \frac{c_2}{\Lambda^2} \epsilon_{abcd} \,(\hat{Q}^a_L \epsilon \,\hat{Q}^b_L) (\hat{Q}^c_L \epsilon \,\hat{Q}^d_L) \ , \ \ \ \ \  \ \ 
\eea
violate baryon and dark matter number by one unit, and the initially produced  asymmetries are related via
$\Delta B_i = -\Delta \tilde{u}'_1$. The baryon asymmetry is then depleted because of the effect of electroweak sphalerons to
\bea
|\Delta B| = \frac{28}{79}|\Delta \tilde{u}'_1| \ .
\eea
The mass of the $\tilde{u}'_1$ particle in such an asymmetric dark matter scenario is
\bea\label{m2}
m_{ \tilde{u}'_1} \approx \frac{\Omega_{ \tilde{u}'_1}}{\Omega_B} \frac{|\Delta B|}{|\Delta \tilde{u}'_1 |} \,m_{\rm proton} \ .
\eea
If all of the dark matter consists of the $\tilde{u}'_1$ particles, then their mass is $m_{ \tilde{u}'_1} \approx 1.75 \ {\rm GeV}$. However, similarly as before, this mass can be smaller if the $\tilde{\nu}'_1$ particles also contribute.

With  two  dark matter candidates, the dark matter relic abundance is given by the sum of the two contributions
\bea
\Omega_{\rm DM} = \varepsilon \, \Omega_{ \tilde{\nu}'_1} + (1-\varepsilon) \, \Omega_{ \tilde{u}'_1} \ ,
\eea
where $\varepsilon$ can take any value between 0 and 1. Given the relations in Eqs.\,(\ref{m1}) and (\ref{m2}), one of the particles $\tilde{\nu}'_1$ and $\tilde{u}'_1$ can have a mass smaller than that of  the neutron, introducing a possible connection to models proposed to explain the neutron lifetime puzzle \cite{Fornal:2018eol}.

\section{Domain wall signatures}
\label{s4}

When the $\rm SU(2)_\ell$ symmetry is spontaneously broken, patches of the Universe undergo a first order phase transition to one of the vacua described earlier: $\vec\psi_{\rm vac1}$ or $\vec\psi_{\rm vac2}$. 
Since those two vacua correspond to disconnected manifolds,  domain walls are create along their boundaries. The breaking of the $\mathcal{Z}_2$ symmetry between the vacua is essential, since without it domain walls would remain stable, leading to cosmological problems, as they would result in unacceptably 
large density fluctuations in the Universe
 \cite{Saikawa:2017hiv}. 

In our model the $\mathcal{Z}_2$ symmetry is  softly broken by the small
 $m_{12}^2$ term, so that  the vacua $\vec\psi_{\rm vac1}$ and $\vec\psi_{\rm vac2}$ have slightly different energy densities. This introduces an instability of  the created domain walls and leads to their annihilation, which in turn gives rise to a stochastic gravitational wave background. It is worth noting that  for domain walls to actually form, the $\mathcal{Z}_2$ symmetry can only be softly broken, otherwise  patches of the Universe would predominantly tunnel to the lower energy density state and no topological defects would arise.

\subsection{Domain wall creation and annihilation}

Choosing the $z$-axis to be perpendicular to the domain wall, the profile of the static configuration $\vec\psi_{dw}(z)$  is given by the solution of the equation 
\bea
\frac{d^2 \vec\psi_{dw}(z)}{d z^2} - \vec\nabla_{\!\psi}V_{\rm eff}\big[\vec\psi_{dw}(z)\big]= 0 \ ,
\eea
subject to the boundary conditions:
 \bea 
 \vec\psi_{dw}(z=-\infty) = \vec\psi_{\rm vac1} \ , \ \ \ \  \ \vec\psi_{dw}(z=\infty) = \vec\psi_{\rm vac2} \ . \ \ \ \ \ \ \ 
 \eea

One of the domain wall parameters, which the resulting gravitational wave signal depends on, is the tension $\sigma$,
\bea
\sigma = \int_{-\infty}^\infty dz \,\rho_{dw}(z) \ ,
\eea
 where $\rho_{dw}$ is the energy density of the domain wall,
\bea
\rho_{dw}(z) = \frac12 \bigg(\frac{d\vec\psi_{dw}(z)}{d z}\bigg)^2 + V_{\rm eff}\big[\vec\psi_{dw}(z)\big] \ .
\eea
In our case the tension can be estimated as
\bea
\sigma \,\sim \,v_\Psi^3 \ .
\eea

\begin{figure}[t!]
\includegraphics[trim={2.4cm 0.8cm 1cm 0cm},clip,width=9.5cm]{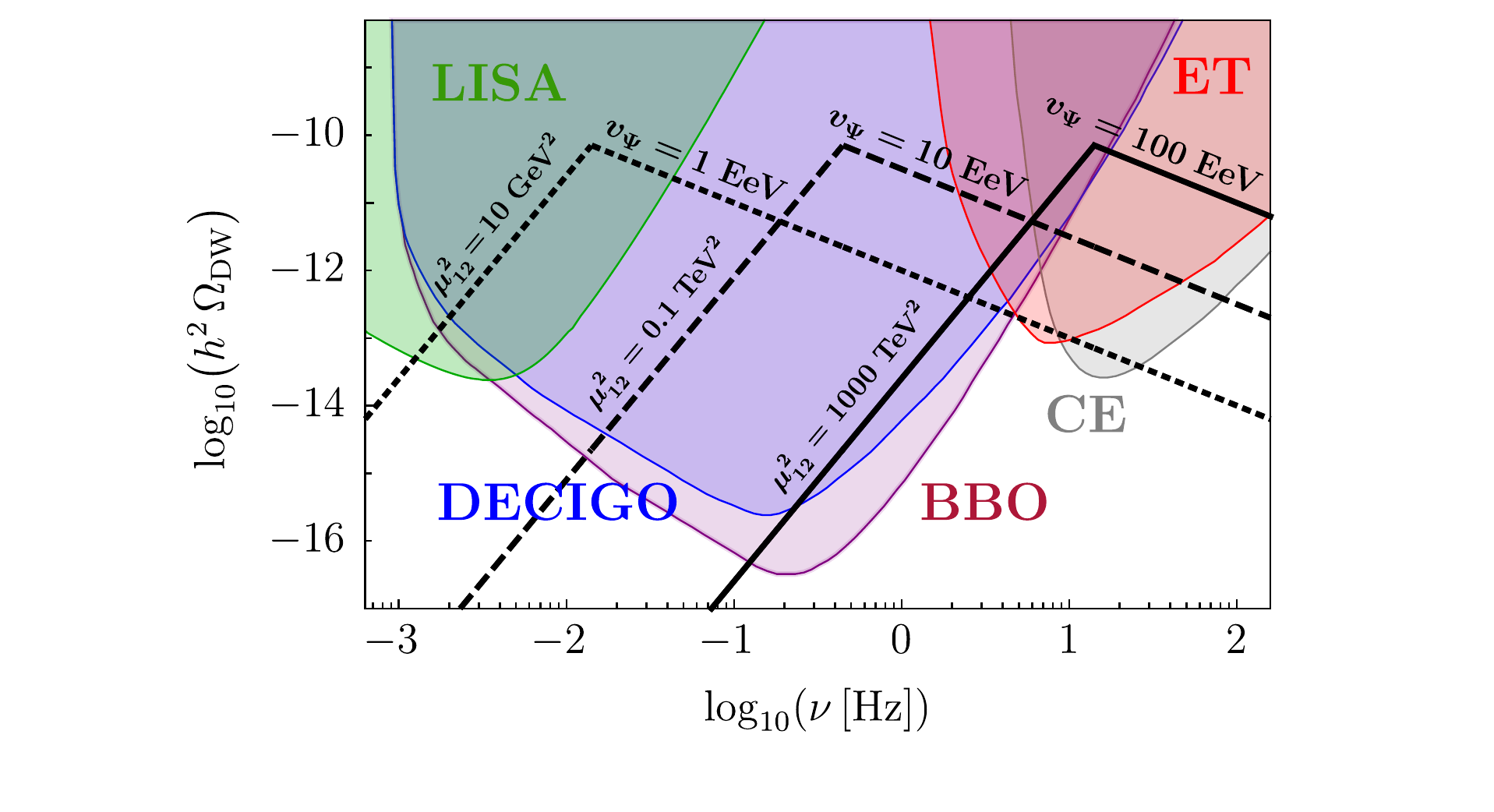} \vspace{-6mm}
\caption{Gravitational wave signals  from domain wall annihilation for several choices of parameters. The colored regions correspond to the reach of future gravitational wave experiments: LISA (green), Big Bang Observer (purple), DECIGO (blue), Einstein Telescope (red), and Cosmic Explorer (gray).}\label{dmwal}
\end{figure} 

The second parameter governing the gravitational wave spectrum is the difference between the energy densities of the  two vacua $\Delta \rho$ (also called the potential bias), which in our model can be approximated by
\bea\label{bias}
\Delta \rho \sim \mu_{12}^2 v_\Psi^2  \ .
\eea
If the potential bias in nonzero, the  domain walls become  unstable and  annihilate when  the volume pressure,
$
p_V \sim \Delta \rho$, exceeds the
pressure due to tension,
$
p_T \sim {\sigma^2}/{M_{Pl}^2}$, where $M_{Pl}$ is the Planck mass.
This  implies the following condition on the parameters of our model,
\bea\label{annn}
\mu^2_{12} \gtrsim \frac{v_\Psi^4}{M_{Pl}^2}\ .
\eea
For example, if $v_\Psi \sim 100 \ {\rm EeV}$ then for the domain walls to annihilate promptly one requires $\mu_{12}^2 \gtrsim 1 \ {\rm TeV^2}$, whereas for $v_\Psi \sim 1 \ {\rm EeV}$ this bound is relaxed to  $10^{-2} \ {\rm GeV^2}$.
The additional constraint on $\mu_{12}^2$ assuring that
domain wall annihilation happens before Big Bang nucleosynthesis (so that it  does not alter the ratios of the produced elements) is much weaker.

\subsection{Gravitational wave signal}

The annihilation of domain walls gives rise to a
 stochastic gravitational wave background described by  \cite{Kadota:2015dza,Saikawa:2017hiv},
\bea\label{dme}
h^2 \Omega_{\rm DW}(\nu)&\approx& 7\times 10^{-21}\left(\frac{g_*}{100}\right)^{-\frac13}\left(\frac{\sigma}{\rm EeV^3}\right)^4\left(\frac{\rm PeV^4}{\Delta \rho}\right)^2\nn\\
&\times& \!\!\left[\left(\frac{\nu}{\nu_d}\right)^3\theta(\nu_d-\nu) + \left(\frac{\nu_d}{\nu}\right)\theta(\nu-\nu_d)\right] , \ \ \ \ \ \ \ 
\eea
where we adopted the values of the area parameter $\mathcal{A} = 0.8$ and efficiency parameter $\tilde\epsilon_{\rm gw} = 0.7$ \cite{Hiramatsu:2013qaa}, $g_*$ is the number of degrees of freedom,
$\theta(x)$ is the Heaviside step function, and $\nu_d$ is the peak frequency given by
\bea
\nu_d \approx (4.5 \ {\rm Hz})\,  \sqrt{\frac{\rm EeV^3}{\sigma}\frac{\Delta \rho}{\rm PeV^4}} \ .
\eea
As described by Eq.\,(\ref{dme}), the spectrum scales like $\sim \nu^3$ for frequencies below the peak frequency, and $\sim 1/\nu$ for higher frequencies.
The
constraints arising from the cosmic microwave background measurements impose the condition
$
h^2 \Omega(\nu) < 2.9 \times 10^{-7}
$ \cite{Clarke:2020bil}, which requires
the parameters $v_\Psi$ and $\mu_{12}^2$ in our model to satisfy the relation
\bea\label{cmb2}
\frac{v_\Psi^4}{\mu^2_{12}} \lesssim (2.5\times 10^9 \ \rm EeV)^2 \ .
\eea

\begin{figure}[t!]
\includegraphics[trim={2cm 0.8cm 1cm 0cm},clip,width=9.0cm]{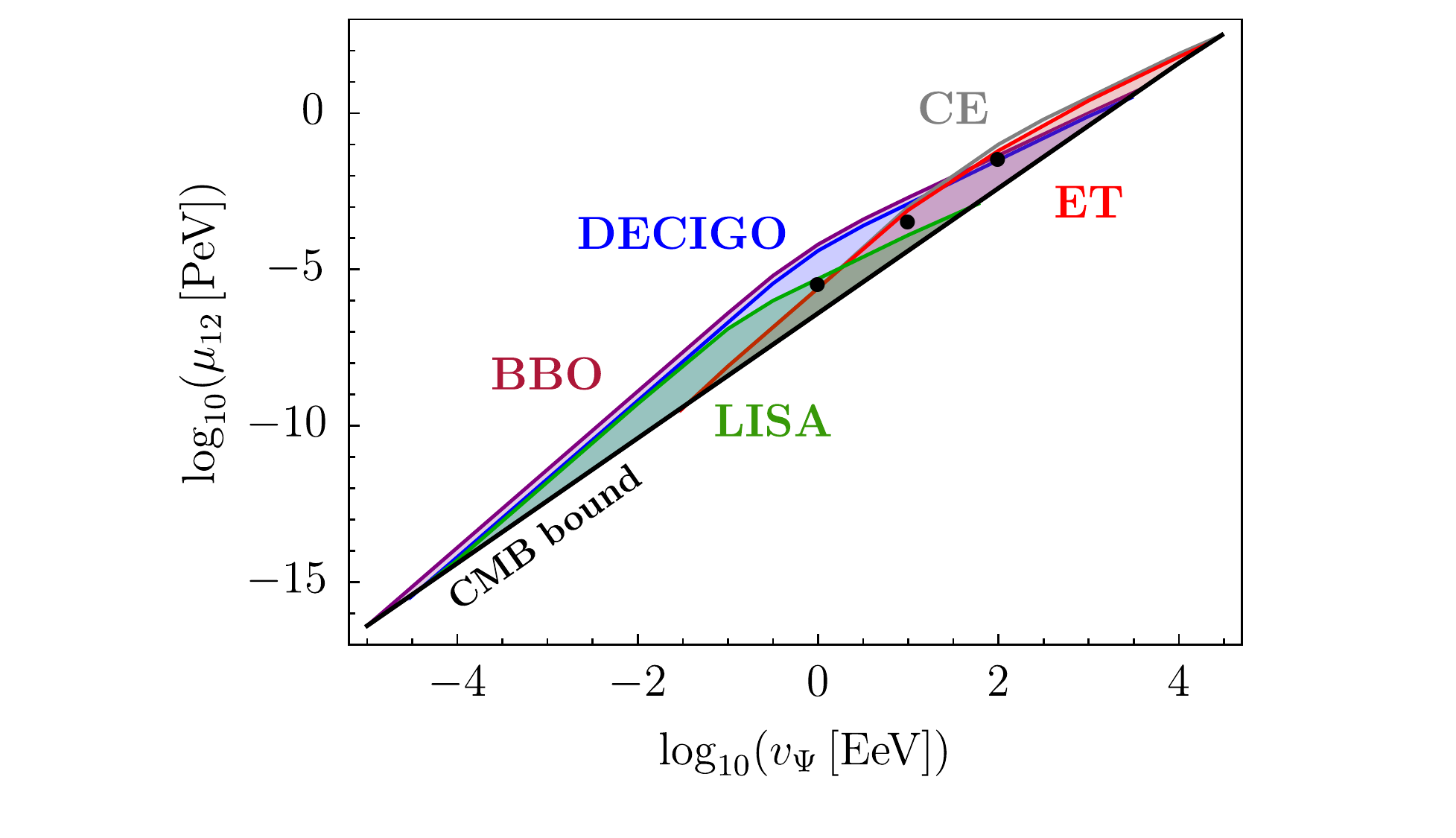} \vspace{-5mm}
\caption{Parameter space regions where the gravitational wave signal from domain walls has a signal-to-noise ratio of at least five after one year of collecting data for various experiments. The black dots correspond to the three signatures shown in Fig.\,\ref{dmwal}.}\label{dwpar}
\end{figure}

Figure \ref{dmwal} shows the expected gravitational wave signal from domain walls for several choices of model parameters, plotted over the sensitivity regions of future gravitational wave detectors: Laser Interferometer Space Antenna  (LISA) \cite{Audley:2017drz},  Big Bang Observer (BBO) \cite{Crowder:2005nr}, DECIGO \cite{Kawamura:2011zz}, Einstein Telescope (ET) \cite{Punturo:2010zz} and Cosmic Explorer (CE) \cite{Reitze:2019iox}. Figure \ref{dwpar} displays regions of parameter space for which the signal can be detected at those experiments -- the upper bound on $\mu_{12}$ is detector-specific (the colors correspond to the selection made in Fig.\,\ref{dmwal}), whereas the lower bound arises from the cosmic microwave background radiation constraint in Eq.\,(\ref{cmb2}).

\section{Phase transition signatures}
\label{s5}

We consider now the possibility of having a first order phase transition associated with ${\rm SU}(4)$ breaking occurring at the PeV scale and producing a measurable gravitational wave signal. 
We first calculate the effective potential of the model. We then proceed to computing the parameters governing the dynamics of the bubble nucleation, and ultimately determine the shape of the possible gravitational wave signatures. 

\subsection{Effective potential}

Denoting the background field by $\phi$, the three contributions to the effective potential are: the tree-level term $V_{0}(\phi)$, the Coleman-Weinberg one-loop correction $V_{\rm CW}(\phi)$, and the finite temperature part $V_T(\phi,T)$, so that
\bea
V_{\rm eff} (\phi, T) = V_{0}(\phi) + V_{\rm CW}(\phi) + V_T(\phi,T)\ .
\eea
Substituting in Eq.\,(\ref{sc2}) the value of $m_\Phi^2$ obtained from minimizing the potential, one obtains
\bea\label{sc2new}
V_0(\phi) = -\frac12 \lambda_\Phi v^2_\Phi \phi^2 + \frac14 \lambda_\Phi \phi^4\ .
\eea
For the Coleman-Weinberg contribution, we use the cutoff regularization scheme and set the minimum of the zero temperature potential and the mass of $\phi$ to be equal to their tree-level values, which results in \cite{Anderson:1991zb}
\bea
V_{\rm CW}(\phi) &=& \sum_{i}\frac{n_i}{(8\pi)^2}\bigg\{m_i^4(\phi) \left[\log \left(\frac{m_i^2(\phi)}{m_i^2(v_\Phi)} \right)-\frac32\right]\nn\\
&&+ \ 2 \,m_i^2(\phi) \,m_i^2(v_\Phi) \bigg\} \ ,
\eea
where contributions from all particles charged under $\rm SU(4)$ are summed over, including $\chi$ (the Goldstone bosons), $n_i$ is the number of degrees of freedom, and $m_i$ are the  background field-dependent masses (we substitute $m_{\chi}(v) \to m_{\Phi}(v)$ for the Goldstone bosons, where $\Phi$ is the radial mode). 
Those masses for the gauge bosons are
\bea
m_{G'}(\phi) &=& \tfrac12 g_4 \phi\ , \nn\\
m_{Z'}(\phi) &=& \tfrac12 \sqrt{\tfrac32 g_4^2 + g_X^2} \, \phi \ ,
\eea
and for the radial mode of the  scalar $\hat\Phi$,
\bea
m_{\Phi} = \sqrt{2 \lambda_\Phi} \, \phi \ .
\eea
The corresponding numbers of degrees of freedom are:\break $n_{G'} = 18$, $n_{Z'} = 3$, $n_\Phi = 1$, and $n_\chi = 7$.

Finally, the finite temperature part of the effective potential is given by \cite{Quiros:2007zz}
\bea
V_T(\phi,T) &=& \frac{T^4}{2\pi^2} \sum_i n_i \int_0^\infty dx\,x^2 \log\left[1\mp e^{-\sqrt{x^2+\tfrac{m_i^2(\phi)}{T^2}}}\right]\nn\\
&&  \hspace{-1.1cm} + \,\frac{T}{12\pi} \sum_k n'_k \left\{m_k^3(\phi)-\left[m_k^2(\phi) + \Pi_k(T)\right]^{3/2}\right\} ,
\eea
where the first line is generated by one-loop diagrams (the sum is over all particles), while the second line arises from the Daisy diagrams (the sum is over bosons only). The prime symbol indicates that only longitudinal degrees of freedom in the case of vector bosons  are included (i.e., $n'_{G'} = 6$, $n'_{Z'} = 1$, $n'_\Phi = 1$, $n'_\chi = 7$), and $\Pi_k(T)$ are the thermal masses \cite{Comelli:1996vm}, which  in our model  are calculated to be (for $\lambda_\Phi \ll 1$)
\bea
&&\Pi_{G'}(T) = \Pi_{Z'}(T) =\tfrac{8}{3} g_4^2 \,T^2 \ ,\nn\\
&&\Pi_{\Phi}(T) = \Pi_{\chi}(T) = \tfrac1{16}\!\left(g_X^2 +\tfrac{15}{2}g_4^2\right)T^2 \ .
\eea

\begin{figure}[t!]
\includegraphics[trim={1.5cm 0.8cm 1cm 0cm},clip,width=8.8cm]{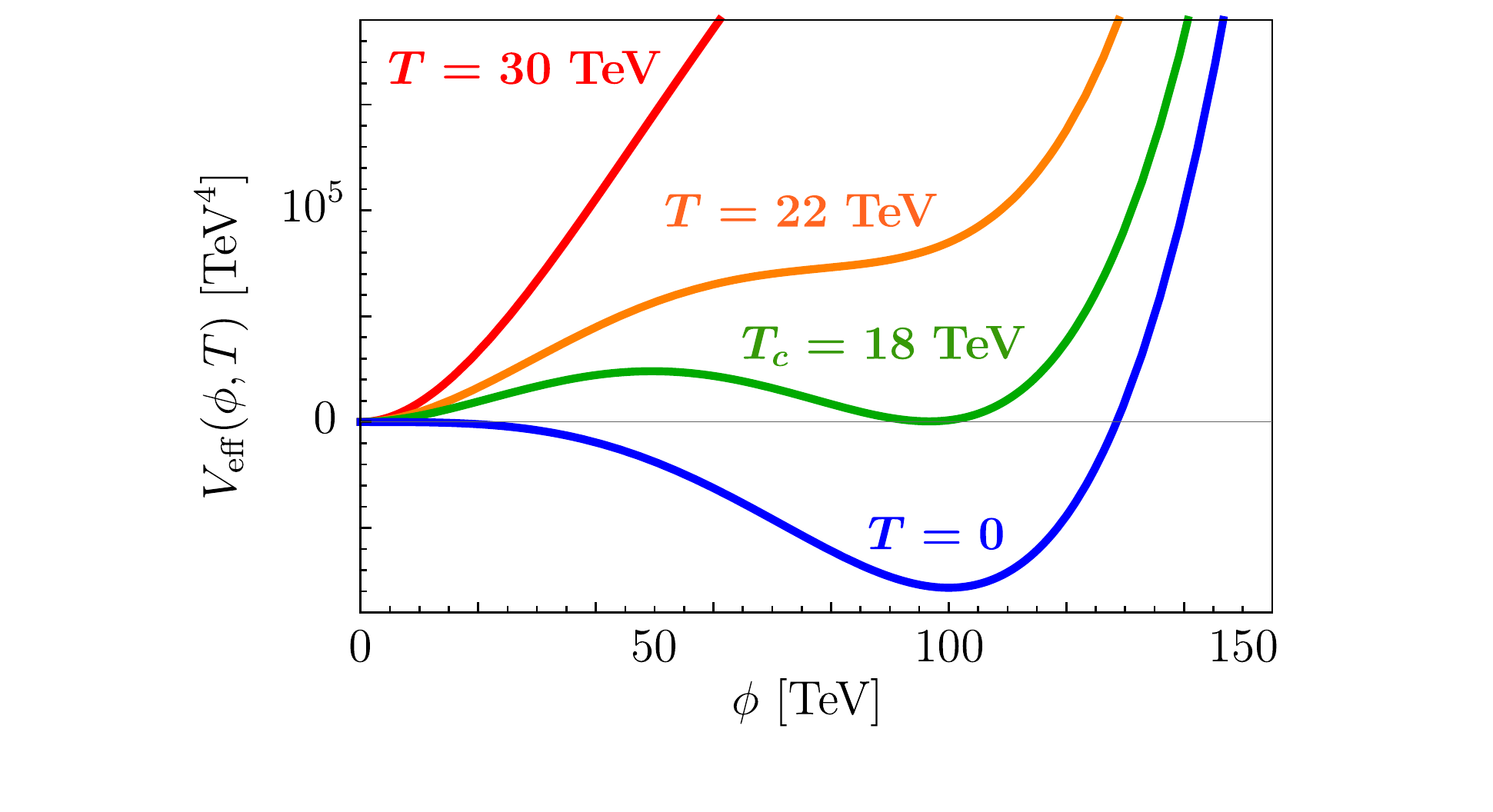} \vspace{-6mm}
\caption{Effective potential of the $\rm SU(4)$ sector of the  model,  $V_{\rm eff}(\phi, T)$, assuming  $v_\Phi = 100 \ \rm TeV$ and  $\lambda_\Phi = 0.004$,  for several choices of temperature.}\label{fig:5}
\end{figure}

 Figure \ref{fig:5} shows the resulting  effective potential plotted adopting $v_\Phi = 100 \ \rm TeV$,  the values of $g_4$ and $g_X$ at that scale (as  discussed in Sec.\,\ref{subb}), and $\lambda_\Phi = 0.004$.  With decreasing temperature, a new true vacuum is formed with a lower energy density than the high temperature false vacuum. Since there is  a potential barrier between the two minima, a  first order phase transition becomes possible.

\subsection{Bubble nucleation}

When the temperature drops below the so-called nucleation temperature $T_*$, this initiates the transition of various parts of the Universe  from the false vacuum to the true one. Such first order phase transitions correspond to the nucleation of bubbles of true vacuum. The nucleation temperature can be determined from the condition that  the bubble nucleation rate  $\Gamma(T)$ becomes comparable to the Hubble expansion rate,  
\bea\label{gam}
\Gamma(T_*) \sim H(T_*)^4 \ . 
\eea
The nucleation  rate is given by  \cite{LINDE1983421}
\bea\label{PT_Gam}
\Gamma(T) \approx \left(\frac{S(T)}{2\pi T}\right)^{\frac32}T^4 e^{-{S(T)}/{T}} \ ,
\eea
in which $S(T)$ is  the Euclidean action calculated as
\bea
S(T)= 4\pi\int dr\,r^2 \left[\frac12\left(\frac{d\phi_b}{dr}\right)^2+V_{\rm eff}(\phi_b, T)\right] , \ \ \ \ \ 
\eea
where $\phi_b(r)$  is the bounce solution describing the profile of the expanding bubble, i.e., the solution to the equation
\bea
\frac{d^2 \phi}{dr^2}+\frac{2}{r}\frac{d\phi}{dr}- \frac{d}{d\phi}V_{\rm eff}(\phi,T)= 0 \ ,
\eea
subject to the following boundary conditions:
\bea
\frac{d\phi}{dr}\bigg|_{r=0} = 0 \ , \ \ \ \ \ \phi(\infty) = \phi_{\rm false} \ .
\eea
The nucleation temperature is calculated, using Eqs.\,(\ref{gam}) and (\ref{PT_Gam}), as the solution to the equation
\bea\label{nucl_temp}
\frac{S(T_*)}{T_*}   \approx  4\log\!\left[\frac{M_{Pl}}{T_*}\right]  - 2\log\left[\frac{4\pi^3g_*}{45}\!\left(\frac{2\pi \,T_*}{S(T_*)}\right)^{\!\!\frac34}\right]. \ \ \ \ \ \ 
\eea

\subsection{Gravitational wave signal}

During the nucleation and the violent expansion of bubbles of true vacuum, a stochastic gravitational wave background  is generated via bubble wall collisions, sound shock waves  in the primordial plasma, and magnetohydrodynamic turbulence. The  shape of the spectrum is determined through simulations and depends only on four quantities: bubble wall velocity $v_w$ (assumed here to be equal to the speed of light; see \cite{Espinosa:2010hh,Caprini:2015zlo} for more details),  nucleation temperature $T_*$, strength of the phase transition $\alpha$, and duration of the phase transition $1/\tilde\beta$. Those parameters depend on the shape of the effective potential, which itself is specific to the particular particle physics model under investigation. This introduces a correspondence between the fundamental parameters of the Lagrangian and the resulting  gravitational wave signal. 

Upon  determining the nucleation temperature $T_*$ by solving Eq.\,(\ref{nucl_temp}), the parameter $\alpha$ is calculated as 
\bea\label{alpha}
\alpha = \frac{\rho_{\rm vac}(T_*)}{\rho_{\rm rad}(T_*)} \ ,
\eea
i.e., as the ratio of the difference between the energy densities of the true and false vacua,
\bea
\rho_{\rm vac}(T) &=& V_{\rm eff}(\phi_{\rm false},T) -  V_{\rm eff}(\phi_{\rm true},T)\nn\\
&-&T \frac{\partial}{\partial T} {\Big[ V_{\rm eff}(\phi_{\rm false},T) -  V_{\rm eff}(\phi_{\rm true},T)\Big]}, \ \ \ \ \ \ 
\eea
and the radiation energy density
\bea
\rho_{\rm rad}(T) = \frac{\pi^2}{30} g_* T^4 \ .
\eea
The parameter $\tilde\beta$ is computed as
\bea\label{beta}
\tilde{\beta} = T_* \frac{d}{dT} \!\left[\frac{S(T)}{T}\right]\bigg|_{T=T_*} \ .
\eea

Based on numerical simulations, the contribution to the stochastic gravitational wave spectrum from sound waves is given by the empirical formula \cite{Hindmarsh:2013xza,Caprini:2015zlo}
\bea\label{swav}
\hspace{-3mm}h^2 \Omega_{\rm sound}(\nu) \,&\approx&\, \frac{1.9 \times 10^{-5}}{\tilde\beta}\left(\frac{g_*}{100}\right)^{\!-\frac13}\left(\frac{\alpha\,\kappa_s}{\alpha+1}\right)^2\nn\\
&\times&\frac{(\nu/\nu_s)^3}{\big[1+0.75\, (\nu/\nu_s)^2\big]^{\frac72}} \  \Upsilon \ ,
\eea
where the fraction of the latent heat transformed into the bulk motion of the plasma \cite{Espinosa:2010hh} and the peak frequency  are
\bea\label{new_swf}
\kappa_s &=& \frac{\alpha}{0.73+0.083\sqrt\alpha + \alpha} \ ,\nn\\
\nu_s &=& (0.019 \ {\rm Hz} ) \left(\frac{T_*}{100 \ {\rm TeV}}\right)\left(\frac{g_*}{100}\right)^\frac16  \tilde\beta \ ,\ \ \ \ \ \ \ 
\eea
and the suppression factor is given by \cite{Ellis:2020awk,Guo:2020grp}
\bea\label{ups}
\Upsilon = 1- \frac1{\sqrt{1+\frac{8\pi^{1/3}}{\tilde{\beta}}\sqrt{\frac{\alpha+1}{3\alpha  \kappa_s}}}} \ . \ \ \ \ \ 
\eea

The gravitational wave spectrum contribution  from bubble wall collisions is estimated to be \cite{Kosowsky:1991ua,Huber:2008hg,Caprini:2015zlo}
\bea\label{col}
h^2 \Omega_{\rm collision}(\nu) \,&\approx&\, \frac{4.9\times 10^{-6}}{\tilde\beta^2}\left(\frac{\alpha\,\kappa_c}{\alpha+1}\right)^2\left(\frac{g_*}{100}\right)^{\!-\frac13}\nn\\
&\times&\frac{(\nu/\nu_c)^{2.8}}{1+2.8\, (\nu/\nu_c)^{3.8}} \ ,
\eea
where the fraction of the latent heat deposited into the bubble front \cite{Kamionkowski:1993fg} and the peak frequency are
\bea
\kappa_c &=& \frac{\frac{4}{27}\sqrt{\frac{3}{2}\alpha} + 0.72\,\alpha }{1+0.72\,\alpha}\ , \nn\\
\nu_c &=& (0.0037 \ {\rm Hz} ) \left(\frac{T_*}{100 \ {\rm TeV}}\right)\left(\frac{g_{*}}{100}\right)^\frac16\tilde\beta \ . \ \ \ \ \ \  
\eea

Finally, magnetohydrodynamic turbulence adds the following contribution  \cite{Caprini:2006jb,Caprini:2009yp},
\bea\label{turb}
h^2 \Omega_{\rm turb}(\nu) \,&\approx&\, \frac{3.4\times 10^{-4}}{\tilde\beta}\left(\frac{\alpha\,\epsilon \, \kappa_s}{\alpha+1}\right)^{\frac32}\left(\frac{g_*}{100}\right)^{-\frac13}\nn\\
&\times& \frac{({\nu}/{\nu_t})^{3}}{\big(1+{8\pi \,\nu}/{\nu_*}\big)\big(1+{\nu}/{\nu_t}\big)^{{11}/{3}}} \ ,
\eea
where  the turbulence suppression  parameter $\epsilon=0.05$ \cite{Caprini:2015zlo}, the peak frequency is
\bea
\nu_t &=& (0.027 \ {\rm Hz} )\left(\frac{T_*}{100 \ {\rm TeV}}\right) \left(\frac{g_*}{100}\right)^\frac16{\tilde\beta} \ ,
\eea
and the parameter $\nu_*$ \cite{Caprini:2015zlo} is given by
\bea\label{lasteq}
\nu_* = (0.017 \ {\rm Hz})\left(\frac{T_*}{100 \ {\rm TeV}}\right) \left(\frac{g_*}{100}\right)^\frac16\ .
\eea
The three contributions add up linearly, resulting in
\bea
h^2 \Omega_{\rm PT} = h^2 \Omega_{\rm sound} + h^2 \Omega_{\rm collision}+ h^2 \Omega_{\rm turb} \ . \ \ \ \ 
\eea

\begin{figure}[t!]
\includegraphics[trim={2.4cm 0.8cm 1cm 0cm},clip,width=9.5cm]{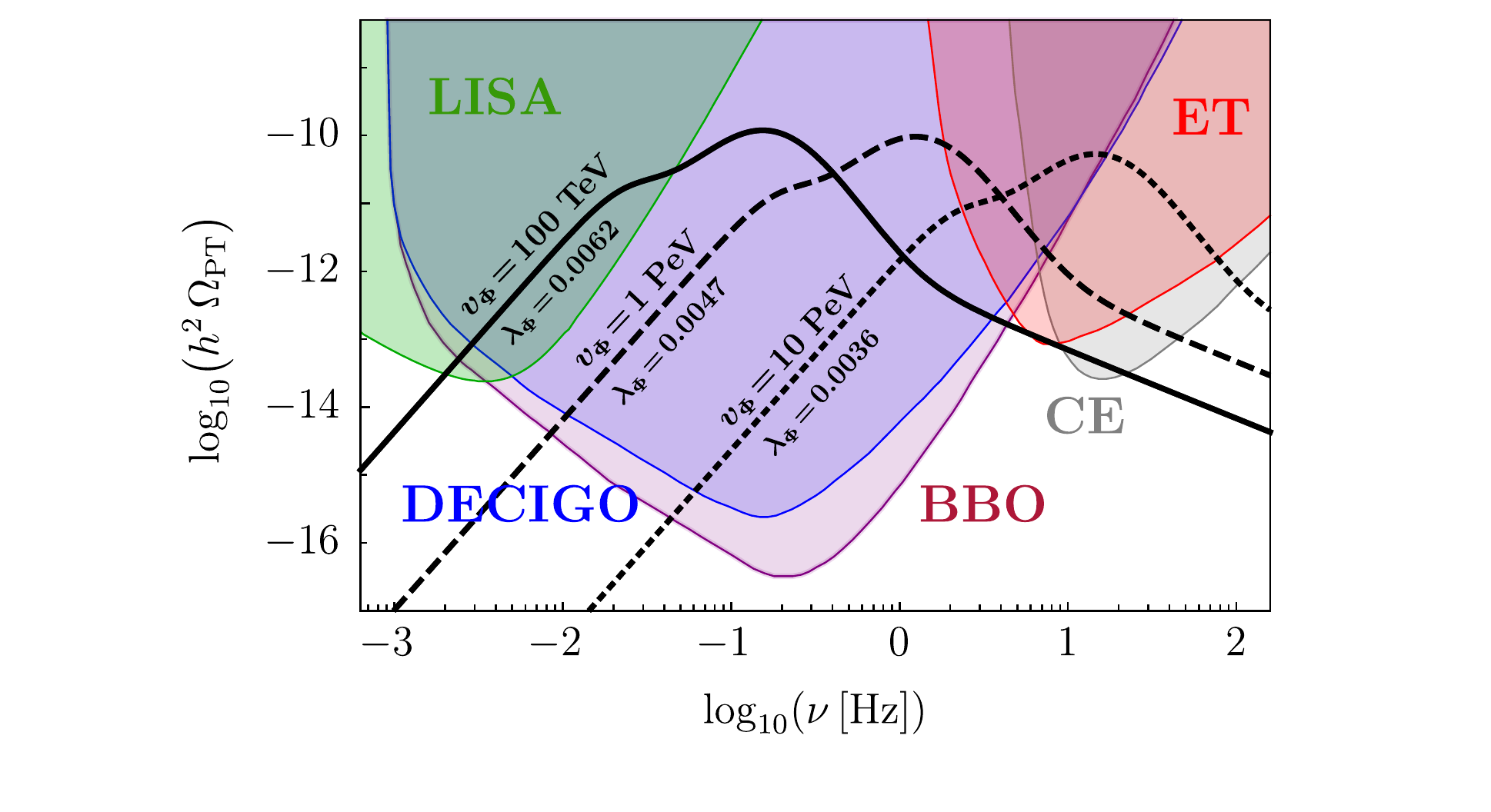} \vspace{-7mm}
\caption{Gravitational wave signals  from a first order phase transition for several choices of parameters. The colored regions correspond to the reach of future gravitational wave experiments,  as in Fig.\,\ref{dmwal}. \\}\label{fig:6}
\end{figure}

\begin{table}[h!] 
\begin{center}
\begingroup
\setlength{\tabcolsep}{6pt} 
\renewcommand{\arraystretch}{1.5} 
\begin{tabular}{ |c |c |c|c|| c| c|c|} 
\hline
 \multicolumn{4}{|c||}{Lagrangian parameters}  & \multicolumn{3}{|c|}{\!Phase transition parameters\!\!} \\ 
\hline
\hline
 \!\!$v_\Phi \, [\rm TeV]$\!\! & $\lambda_\Phi$ &   $g_4$ &   $g_X$ & $\alpha$ & $\tilde\beta$   & $T_*\, [\rm TeV]$ \\ 
\hline
\hline
 $100$   &    $0.0062$&   $0.88$  & $0.30$ &    9 & 220 & $3.2 $ \\[1pt]
\hline
$1000$   &     $0.0047$  &  $0.82$ &   $0.29$& \ \,30\,   \ &  \ 260 \ & $ 22$ \\[1pt]
\hline
$10000$   &     $0.0036$  &  $0.77$ &   $0.29$& 43  &  \ 350 \ & $ 190$ \\[1pt]
\hline
\end{tabular}
\endgroup
\end{center}
\vspace{-3mm}
\caption{Phase transition and fundamental Lagrangian parameters for the three  gravitational wave signatures presented in Fig.\,\ref{fig:6}.}\vspace{1mm}
\label{tabletransl}
\end{table}

Figure \ref{fig:6} illustrates the gravitational wave signatures of the model arising from a first order phase transition in the early Universe  for three different $\rm SU(4)$ symmetry breaking scales: $v_\Phi = 100 \ \rm TeV$ (solid line), $1 \ \rm PeV$ (dashed line), and $10 \ \rm PeV$ (dotted line),  plotted for the  quartic couplings $\lambda_\Phi$ which amplify the signal. In all of these cases the three contributions to the gravitational wave spectrum are visible: sound shock wave (main peak), bubble collision (to the left of the peak frequency), and magnetohydrodynamic turbulence (to the right of the peak). This is the result of the suppression factor $\Upsilon$ in Eq.\,(\ref{swav}), which reduces the contribution from sound waves roughly by a factor of 100 -- without this suppression the gravitational wave signal would be dominated entirely by the sound wave component.

The translation between the fundamental Lagrangian parameters $(v_\Phi, \lambda_\Phi, g_4, g_Y)$ and the phase transition parameters $(T_*, \alpha, \tilde\beta)$ for each of the expected signals shown in Fig.\,\ref{fig:6} is provided in Table \ref{tabletransl}.  The gauge couplings $g_4$ and $g_Y$ are fixed by the running of the Standard Model strong and electroweak couplings to have particular values at a given symmetry breaking scale, thus the only free fundamental parameters are $v_\Phi$ and $\lambda_\Phi$. The variation in the  shape of the spectra shown in Fig.\,\ref{fig:6} arises precisely from the fact that the gauge couplings vary depending on the symmetry breaking scale --  this affects the shape of the effective potential, thus different  values of the quartic coupling $\lambda_\Phi$ are required to amplify the signal.

Phase transitions corresponding to a higher  symmetry breaking scale  are characterized by a larger nucleation temperature, which  shifts the signal toward higher frequencies. 
A shift in the same direction occurs for transitions described by a larger parameter $\tilde\beta$. The height of the peak is determined by both the strength of the phase transition  $\alpha$ and its duration $1/\tilde\beta$: the signal is stronger for larger values of $\alpha$ and for smaller values of  $\tilde\beta$ (longer phase transitions).

\begin{figure}[t!]
\includegraphics[trim={1.4cm 0.4cm 1cm 0cm},clip,width=9.3cm]{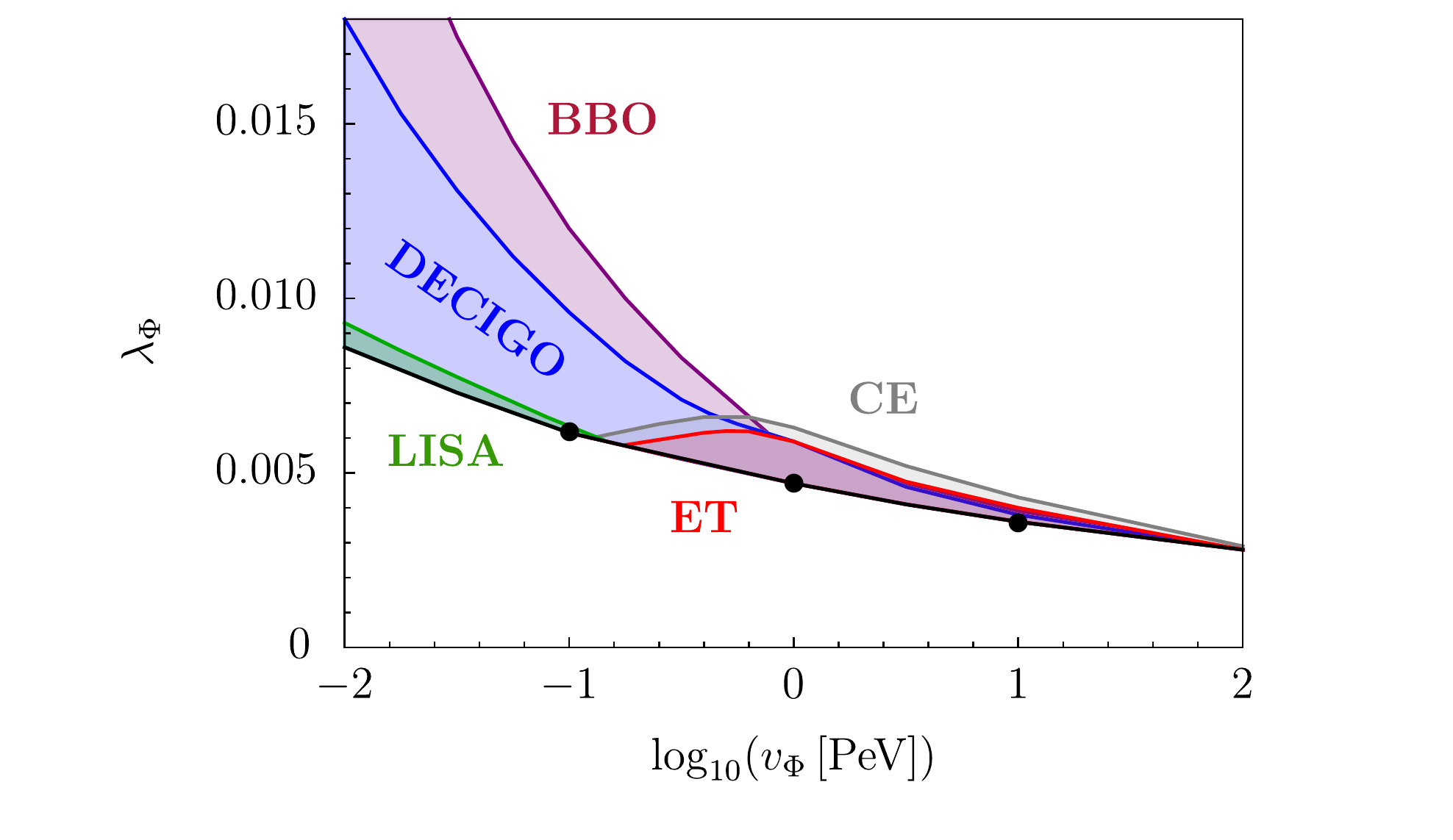} \vspace{-7.7mm}
\caption{Parameter space regions where the gravitational wave signal from a first order phase transition has a signal-to-noise ratio of at least five after one year of collecting data for various experiments. The black dots correspond to the three signatures shown in Fig.\,\ref{fig:6}.}\label{fig:7}
\end{figure}

In addition to the signals themselves, the sensitivities of the future gravitational wave experiments: LISA, DECIGO, Big Bang Observer, Einstein Telescope, and Cosmic Explorer, are also shown in Fig.\,\ref{fig:6}. To investigate in more detail how this reach translates into probing the fundamental parameters in the Lagrangian, we scanned over $(v_\Phi, \lambda_\Phi)$ and determined the  regions for which the above experiments will be sufficiently  sensitive to detect the first order phase transition signals. The results of this scan are presented in Fig.\,\ref{fig:7}.

\section{Novel gravitational wave signature}\label{s6}

An intriguing scenario arises when there is a large hierarchy between the $\rm SU(2)_\ell$ and $\rm SU(4)$ symmetry breaking scales. This offers the possibility of having the domain wall signal and the first order phase transition signal coexist, both being within the sensitivity region of upcoming gravitational wave experiments, and producing novel features to search for in the gravitational wave spectrum.

To realize this unique signature, we assume that the $\rm SU(2)_\ell$ symmetry is broken at the scale $v_\Psi \sim 100 \ \rm EeV$, whereas the $\rm SU(4)$  is broken at $v_\Phi \sim 100 \ \rm TeV$. The breaking of  $\rm SU(2)_\ell$ at the high scale leads to the production of domain walls, which undergo annihilation  (due to the small nonzero $\mu_{12}^2$ term) and produce a gravitational wave signal corresponding to the rightmost curve in Fig.\,\ref{dmwal}. On the other hand, the $\rm SU(4)$ symmetry breaking at the lower scale results in a first order phase transition, which leads to a gravitational wave signal analogous to the leftmost curve in Fig.\,\ref{fig:6}.

\begin{figure}[t!]
\includegraphics[trim={2.4cm 0.8cm 1cm 0cm},clip,width=9.5cm]{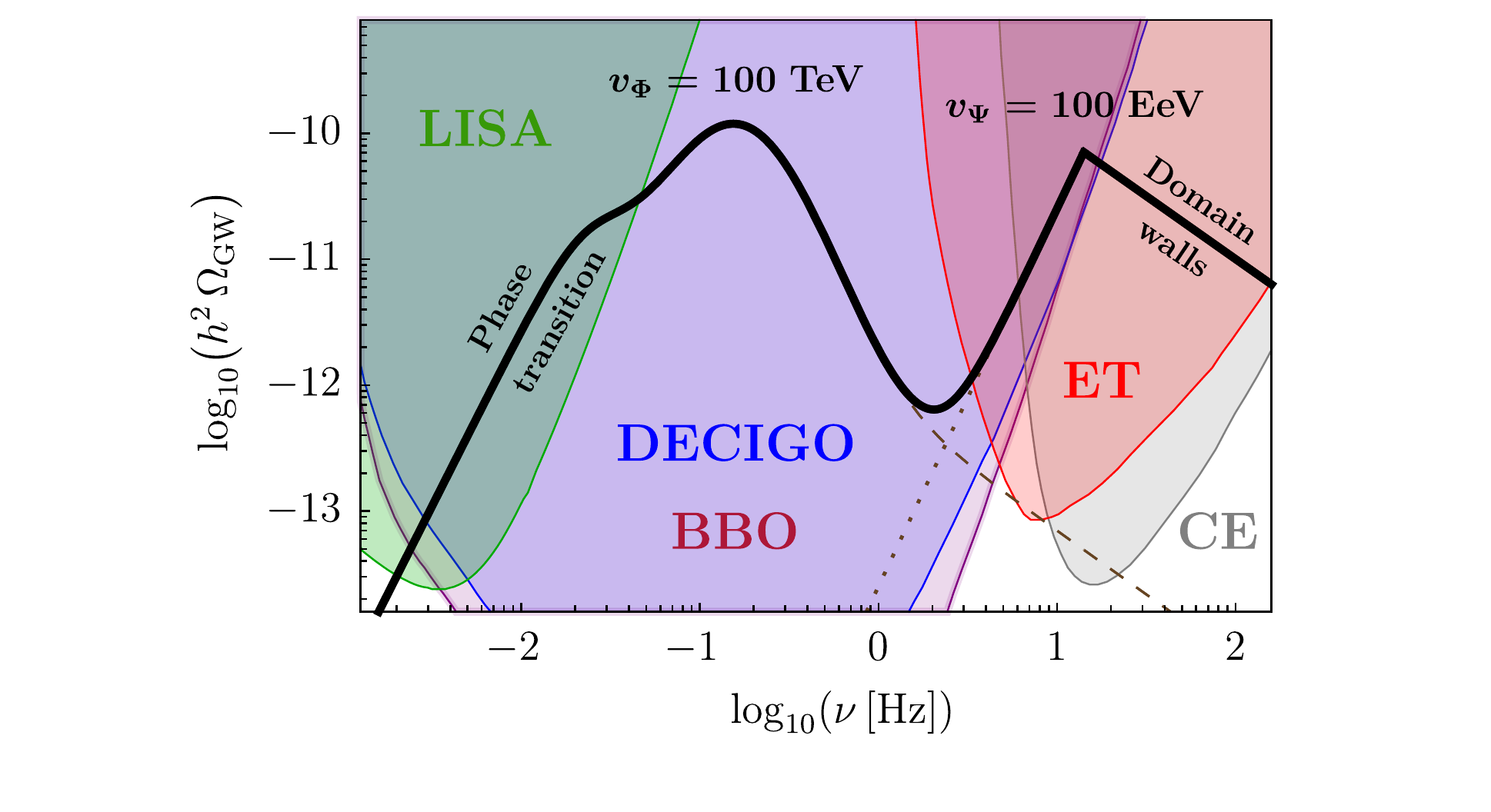} \vspace{-6mm}
\caption{Novel gravitational wave signature of the model expected when a large hierarchy between symmetry breaking scales exists. The signal is searchable in future gravitational wave detectors: LISA, DECIGO, Big Bang Observer, Einstein Telescope, Cosmic Explorer.}\label{fig:8}
\end{figure}

The two contributions combine to form a new double-bump gravitational wave signature shown in Fig.\,\ref{fig:8}. The signal consists of a smooth peak arising from a first order phase transition, and a second sharp peak corresponding to domain wall annihilation. The slopes for the two peaks are different and are described by the frequency power law behavior in Eqs.\,(\ref{dme}), (\ref{swav}), (\ref{col}), and (\ref{turb}), which can be used to differentiate between the two contributions. 

Another promising property of the signal is that it can be searched for throughout a wide range of frequencies, making it relevant for a number of upcoming gravitational wave experiments: LISA, DECIGO, Big Bang Observer, Einstein Telescope, and Cosmic Explorer. Moreover, several parts of the spectrum can be probed by more than one experiment, which would foster collaboration between the different groups in case of a future discovery.

\section{Summary and Outlook}\label{s7}

Physics beyond the Standard Model is certainly necessary to explain several outstanding questions such as: What is dark matter? How did the matter-antimatter asymmetry originate? How do neutrino get their masses? In the time when conventional particle physics experiments, although working at the cutting edge of technological development and gathering extremely  valuable data about the smallest and largest scales in the Universe, 
have not brought us answers to those questions, gravitational wave astronomy has provided us with a glimmer of hope to attack those problems from a  new direction.  

Processes taking place in the early Universe and triggered by new physics, such as first order phase transitions, domain wall annihilation, or the dynamics of cosmic strings, can produce a stochastic gravitational wave background  detectable in current and future experiments, offering us precious insight into the particle physics  at the highest energies. This makes analyzing models leading to such gravitational wave signals a worthwhile endeavor.

In this paper, we propose a particle physics framework which enables answering the aforementioned  open questions and gives rise to a unique gravitational wave signature. The new model contains two dark matter candidates and allows to generate the matter-antimatter asymmetry of the Universe in an asymmetric dark matter setting through two distinct symmetry breaking events. Unlike in the case of typical  asymmetric dark matter models, our scenario does not require the masses of the dark matter particles  to be fixed at particular values.  Depending on the individual contribution of each of them to the dark matter relic density, the mass of one of them can be below the mass of the neutron, opening up a possible connection to models explaining the neutron lifetime anomaly through a dark decay of the neutron \cite{Fornal:2018eol}.

The new signature consists of two peaks: one arising from a first order phase transition triggered by $\rm SU(4)$ symmetry breaking at  $\sim 100 \ \rm TeV$, and the second one resulting from domain wall annihilation after $\rm SU(2)_\ell$ breaking at $\sim 100 \ \rm EeV$.  The signal spans a wide range of frequencies  relevant for the upcoming  gravitational wave experiments:  LISA, DECIGO, Big Bang Observer, Einstein Telescope, and Cosmic Explorer. Since parts of the  predicted signal lie in regions of overlapping sensitivities of various detectors, it also offers an opportunity of cross-checking the results, encouraging stronger collaboration within the gravitational wave physics community.

We note that several interesting variations of the proposed signature are possible. For example, the order of the two peaks may be reversed, which happens if the $\rm SU(2)_\ell$ is broken at  $\sim 1 \ \rm EeV$ while the $\rm SU(4)$  breaking occurs at $\sim  10 \ \rm PeV$. Another intriguing  scenario is when the peak frequencies of the two bumps coincide (when $\rm SU(2)_\ell$  is broken at  $\sim 10 \ \rm EeV$ and $\rm SU(4)$ is broken at $\sim 1 \ \rm PeV$), leading to an unusual dependence on frequency in the signal peak region.

Finally, it would  be interesting to investigate the possibility of an additional cosmic string contribution to the gravitational wave spectrum. Cosmic strings are typically produced when a $\rm U(1)$ gauge symmetry is spontaneously broken \cite{Kibble:1976sj}, and the resulting gravitational wave signal depends on the scale at which this happens \cite{Vilenkin:2000jqa,Gouttenoire:2019kij}. However, it has recently been shown that cosmic strings can also be produced through the breaking of non-Abelian gauge symmetries. For example, if an
 $\rm SU(2)$ gauge group is broken by the vacuum expectation values of two triplet scalars (instead of the two doublets as in the model we considered),  topologically stable strings can form \cite{Qaisar,Hindmarsh:2016lhy}. The interplay between the cosmic string contribution and other gravitational wave signatures will be the subject of an upcoming publication \cite{Barry}.

A discovery of a stochastic gravitational wave background, such as the one discussed in this paper, would introduce a breakthrough in our understanding of the early Universe, shedding light on what happened just a small fraction of a second after the Big Bang.

\subsection*{Acknowledgments}

This research was supported by the National Science Foundation under Grant No. PHY-2213144.

\bibliography{bibliography}

\end{document}